\documentclass[12pt,a4paper]{article}
\usepackage{graphicx}
\usepackage{times}
\title{\bf Feedback from Massive YSOs and Massive Stars}
\author{You-Hua Chu$^1$ and Robert A.\ Gruendl$^1$\\
\vspace{1cm}\\
\normalsize $^1$ Astronomy Department, University of Illinois, Urbana, IL 61801, USA}
\date{\mbox{}}
\begin{document}
\maketitle
\pagestyle{empty}
%
% WE REDEFINE THE plain LaTeX PAGESTYLE !!! 
% THIS PAGESTYLE WILL BE USED FOR THE FIRST PAGE ONLY !
%
\def\bull{\vrule height .9ex width .8ex depth -.1ex}
\makeatletter
\def\ps@plain{\let\@mkboth\gobbletwo
\def\@oddhead{}\def\@oddfoot{\hfil\tiny\bull\quad
``The multi-wavelength view of hot, massive stars''; 39$^{\rm th}$ Li\`ege Int.\ Astroph.\ Coll., 12-16 July 2010 \quad\bull}%
\def\@evenhead{}\let\@evenfoot\@oddfoot}
\makeatother
%
% AND DEFINE OUR MACROS FOR THE REFERENCE LIST
% I.E \beginrefer \refer and \endrefer
%
\def\beginrefer{\section*{References}%
\begin{quotation}\mbox{}\par}
\def\refer#1\par{{\setlength{\parindent}{-\leftmargin}\indent#1\par}}
\def\endrefer{\end{quotation}}
%
% BEGIN THE ABSTRACT CHAPTER WITH \noindent\small, ENCLOSE IT IN A GROUP
% AND BOLDFACE THE TITLE.
%
{\noindent\small{\bf Abstract:} 
Massive stars are powerful sources of radiation, stellar winds, and
supernova explosions.  The radiative and mechanical energies injected
by massive stars into the interstellar medium (ISM) profoundly alter
the structure and evolution of the ISM, which subsequently influences
the star formation and chemical evolution of the host galaxy.
In this review, we will use the Large Magellanic Cloud (LMC) as a
laboratory to showcase effects of energy feedback from massive young
stellar objects (YSOs) and mature stars.  We will also use the Carina
Nebula in the Galaxy to illustrate a multi-wavelength study of feedback 
from massive star.
}
%
% NOW COMES THE MAIN BODY OF THE ARTICLE
%
\section{Introduction}
Massive stars generate large amounts of energy and are thus luminous.
The energy leaves a star mainly in the form of radiation; only a small 
portion of the energy is imparted to stellar wind through line scattering.
For example, an O5 main sequence star has a luminosity of 
$\sim$10$^6$ $L_\odot$, or $\sim$4$\times$10$^{39}$ ergs~s$^{-1}$, while 
the mechanical luminosity of its stellar wind is only
$L_{\rm w}$ $\sim$1.3$\times$10$^{35}$ ergs~s$^{-1}$,
assuming a typical mass loss rate ($\dot M$) of 10$^{-7}$ $M_\odot$ yr$^{-1}$ 
and a wind terminal velocity ($V_{\rm w}$) of 2,000 km~s$^{-1}$. 
As a massive star evolves and traces a nearly horizontal track in the
theoretical HR diagram, its luminosity is nearly constant, but its 
varying effective temperature leads to different mass loss rates and
stellar wind velocities.
For example, a red supergiant has a higher $\dot M$, 
$\sim$10$^{-4}$ $M_\odot$ yr$^{-1}$, but a lower $V_{\rm w}$, 10--50 km~s$^{-1}$,
and its $L_{\rm w}$ is even lower, $\sim$10$^{34}$ ergs~s$^{-1}$.
A Wolf-Rayet (WR) star, on the other hand, has both a high $\dot M$ and
a high $V_{\rm w}$, and thus the highest $L_{\rm w}$, 
10$^{37}$--10$^{38}$ ergs~s$^{-1}$, which is still much lower than 
its luminosity.

At infancy, the radiation of a massive young stellar object (YSO)
can heat and repel the ambient dust, photo-dissociate molecules, and
photo-ionize atoms, while its stellar wind clears out the circumstellar
material and further erodes the placental molecular cloud.
During its adulthood, a massive star's radiation photo-ionizes and heats 
its ambient interstellar medium (ISM) to 10$^4$ K, and its stellar wind 
dynamically interacts with the ISM, blowing bubbles and generating
turbulence.  At the end of its life, a massive star explodes as a 
supernova, releasing $\sim10^{51}$ ergs of kinetic energy into the ISM,
forming a classical supernova remnant (SNR) in a dense medium, or 
merely heating its ambient medium further if it is in a hot low-density
medium such as the interior of a superbubble.

Stellar energy feedback profoundly alters the structure of the ISM by 
producing interstellar shells up to 10$^3$ pc in size, injecting turbulence,
creating multiple phase components with different physical conditions,
and ejecting hot gas into the galactic halo.  As stars are formed from the
ISM, the physical changes of the ISM affect the future generation
of star formation, either dispersing or compressing the ISM to inhibit
or enhance star formation.  The mass loss from massive stars enriches
the ISM and the intergalactic medium, contributing to the chemical
evolution of a galaxy.  

Sites of stellar energy feedback provide excellent laboratories for
us to study a wide range of astrophysical processes, such as shocks,
thermal conduction, turbulence, cosmic ray acceleration, etc.
Observations of stellar energy feedback allow us to better grasp its
ramifications on cosmic evolution.  It is thus important to study 
stellar energy feedback.  Unlike star formation, however, there is no simple
recipe for stellar energy feedback because the ISM surrounding massive
stars has diverse physical conditions, resulting in complex dynamical
interactions.

\section{The Large Magellanic Cloud as a Laboratory}

The actions of energy feedback from massive stars are best observed
in a galaxy where a clear, high-resolution view of both stars and
the ISM for the entire galaxy can be obtained.  The Large Magellanic
Cloud (LMC) provides such an ideal laboratory to study massive stars 
acting on the ISM because of its nearly face-on orientation, small
distance (50 kpc, where 1$''$ corresponds to 0.25 pc), and low 
foreground extinction.

The LMC has been surveyed extensively for both stars and the ISM:
$UBVI$ photometry of bright stars (MCPS, Zaritsky et al.\ 2004),
emission-line survey of ionized gas (MCELS, Smith and the MCELS Team
1999), {\it ROSAT} X-ray mosaic image of the hot (10$^6$ K) ionized
gas (made by S. Snowden), ATCA+Parkes 21-cm line survey of H\,{\sc i}
(Kim et al.\ 2003), {\it Spitzer} near- and mid-IR surveys of stars 
and dust (SAGE, Meixner et al.\ 2006), and CO surveys of molecular 
clouds (NANTEN, Fukui et al.\  2008; MAGMA, Hughes et al.\ 2010). 
These surveys provide a detailed view of the distributions, physical 
conditions, and kinematics of the ISM and the underlying stellar
population in the LMC. 

\begin{figure}[h!]
%\begin{minipage}{13cm}
\centering
\includegraphics[width=13.5cm]{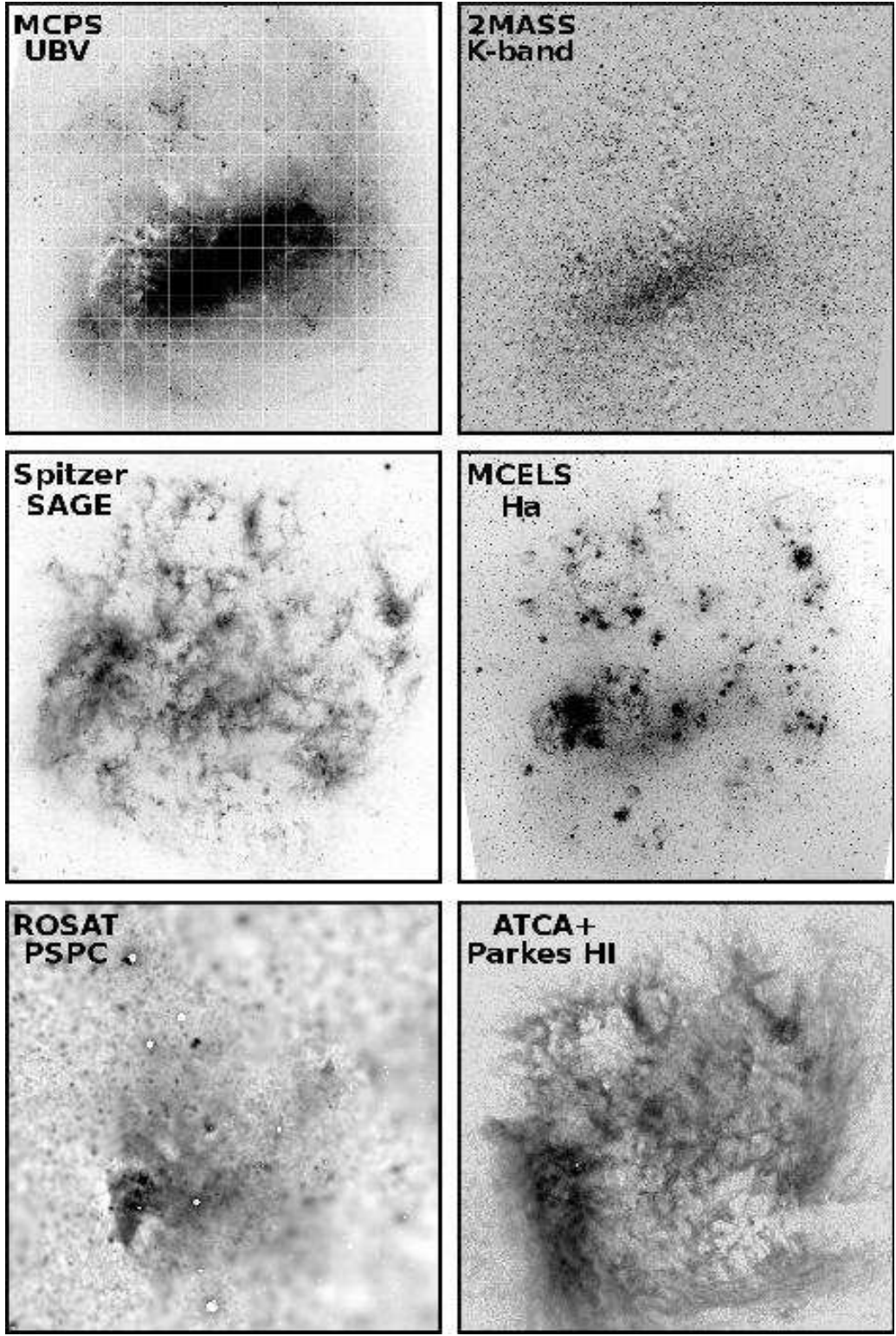}
\caption{Multi-wavelength images of the LMC.  The top row
displays the MCPS composite image from Harris \& Zaritsky (2009) and
2MASS $K$-band image, the middle row the {\it Spitzer} SAGE composite 
image from Meixner et al.\ (2006) and MCELS H$\alpha$ image from 
Smith \& the MCELS Team (1999), and the bottom row the {\it ROSAT} 
PSPC mosaic in the 0.5--2.0 keV band made by S. Snowden and
ATCA+Parkes H\,{\sc i} map from Kim et al.\ (2003).  Note that these
images do not have identical scales and orientation.
\label{fig_1}}
%\end{minipage}
\end{figure}

Figure 1 displays some of the survey images to illustrate 
the full view of the LMC at different wavelengths: MCPS optical
continuum, 2MASS $K$-band, {\it Spitzer} IRAC and MIPS composite,
MCELS H$\alpha$, {\it ROSAT} PSPC mosaic in the 0.5--2.0 keV band,
and ATCA+Parkes H\,{\sc i} 21 cm line.  These detailed surveys of stars
and ISM enable many studies that are not possible in the Galaxy 
or in more distant galaxies.  For example: \begin{itemize}

\item
The MCPS data have been used to determine the spatially-resolved
star formation history of the LMC (Harris \& Zaritsky 2009), and
this star formation history combined with Starburst99 (Leitherer
et al.\ 1999) can be used to estimate the history of stellar energy 
injected into the ISM.  

\item
The actual stellar content of a superbubble can be observed
to estimate the total stellar energy input, and the kinetic
energy in the expanding shell and the thermal energy in the 
superbubble interior can be measured to determine the stellar
energy retained in the ISM.  It is found that the ISM retained
much less energy than the total stellar energy injected
(Cooper et al.\ 2004).

\item
The {\it Spitzer} survey of the LMC can be used to search for
stars with IR excesses indicating circumstellar dust and to identify
YSOs (Gruendl \& Chu 2009).  The power spectrum analysis of 
{\it Spitzer} images show two power laws with different slopes 
joining at a scale of 100--200 pc, which may be a scale height of the
dust disk of the LMC (Puerari et al.\ 2010).

\item 
The stellar mass distribution assessed from the 2MASS survey
and the gas distribution derived from the H\,{\sc i} and CO surveys
can be used to determine the gravitational instability map of
the LMC, and it is found that $\sim$85\% of the massive YSOs
are located within the unstable regions (Yang et al.\ 2007). 

\item
Star formation related to stellar energy feedback can be
studied in detail around OB associations and superbubbles
(Chen et al.\ 2009, 2010) as well as supergiant shells
(Book et al.\ 2009).

\end{itemize}

It is impossible to review every stellar 
energy feedback topic.  In this paper, we have selected a few
obvious topics in which recent progress has been made:
(1) dispersal of molecular clouds, (2) interstellar 
shells, (3) acceleration of cosmic rays, and (4) anatomy
of the Carina Nebula.

\section{Massive Star Formation and Dispersal of Molecular Clouds}

\begin{figure}[t!]
%\begin{minipage}{13cm}
\centering
\includegraphics[width=15.5cm]{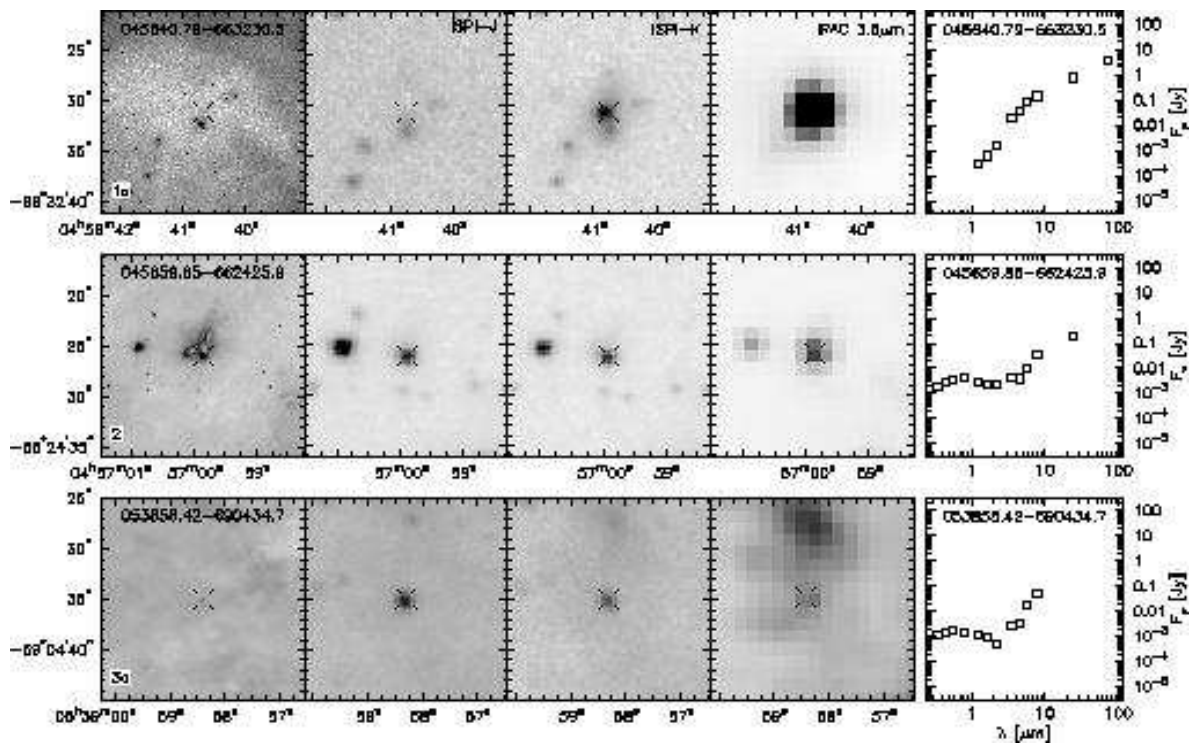}
\caption{Images and SEDs of three YSOs with three different
types of immediate environments: in a dark cloud (top row),
in the tip of a bright-rimmed dust pillar (middle row),
and in a small H\,{\sc ii} region (bottom row).  The images from 
left to right are: {\it HST} H$\alpha$, CTIO 4m ISPI $J$ 
and $K$, {\it Spitzer} IRAC 3.6~$\mu$m.  The rightmost panel 
shows the SED. 
\label{fig_2}}
%\end{minipage}
\end{figure}

To investigate the energy feedback of massive stars in their infancy,
we can examine the immediate surroundings of massive YSOs.
A large sample of massive and intermediate-mass YSOs have been 
identified in the LMC (Gruendl \& Chu 2009).  About 100 of these 
YSOs have archival {\it Hubble Space Telescope} ({\it HST}) H$\alpha$ 
and continuum images available.  These H$\alpha$ images reveal
three types of immediate environments of YSOs: in dark clouds,
inside or on the tip of bright-rimmed dust pillars, and in small 
H\,{\sc ii} regions (Vaidya et al.\ 2009).  Figure 2 shows images 
and spectral energy distributions (SEDs) of three exemplary YSOs.
It is suggested that the three types of environments represent
an evolutionary sequence, as the stellar wind clears out the 
ambient molecular cloud and reveals a small H\,{\sc ii} region.
This evolutionary sequence is supported by the evolutionary 
stages of the YSOs indicated by their SEDs.

The dispersal of ambient molecular material by massive YSOs
is also seen in direct observations of HCO$^+$, a tracer for
dense molecular gas.  Using the ATCA facility, HCO$^+$ has
been mapped in two OB/H\,{\sc ii} complexes: N44 (Seale et al., 
in preparation) and N159 (Chen et al., in preparation).
Many YSOs in these two complexes have  {\it Spitzer} IRS spectra
available.  Based on the progressive presence of silicate absorption, 
PAH emission, and fine-structure atomic line emission, an evolutionary
sequence of YSOs can be defined (Seale et al.\ 2009). It is observed
that the youngest YSOs, those with silicate absorption, are still
coincident with molecular cores indicated by HCO$^+$ peaks, while the 
intermediate-aged YSOs show offsets from molecular cores, and
the most evolved YSOs are no longer associated with molecular cores.
Both the aforementioned {\it HST} H$\alpha$ observations and these HCO$^+$
observations suggest that massive YSOs quickly disperse their ambient
molecular material, probably before they reach the main sequence.

On a larger scale, the dispersal of molecular clouds in the LMC has 
been studied by Kawamura et al.\ (2009).  They find three types of 
molecular clouds: Type I has no massive star formation, as indicated
by a lack of H$\alpha$ emission; Type II has isolated massive star
formation, as indicated by small discrete H\,{\sc ii} regions; 
and Type III has young clusters and prominent H\,{\sc ii} regions 
(see Figure 3).  As 66\% 
of LMC clusters younger than 10 Myr are associated with molecular clouds, 
Type III molecular clouds probably last for $\sim$7 Myr.  If the three types 
of molecular clouds form an evolutionary sequence and the relative population
of these three types is proportional to the time spent in these stages,
then the population ratio of $N_{\rm I}$ : $N_{\rm II}$ : $N_{\rm III}$ = 1: 2 :1
implies that the dispersal timescale of molecular clouds is $\sim$30 Myr.

\begin{figure}[t!]
%\begin{minipage}{13cm}
\centering
\includegraphics[width=8.5cm]{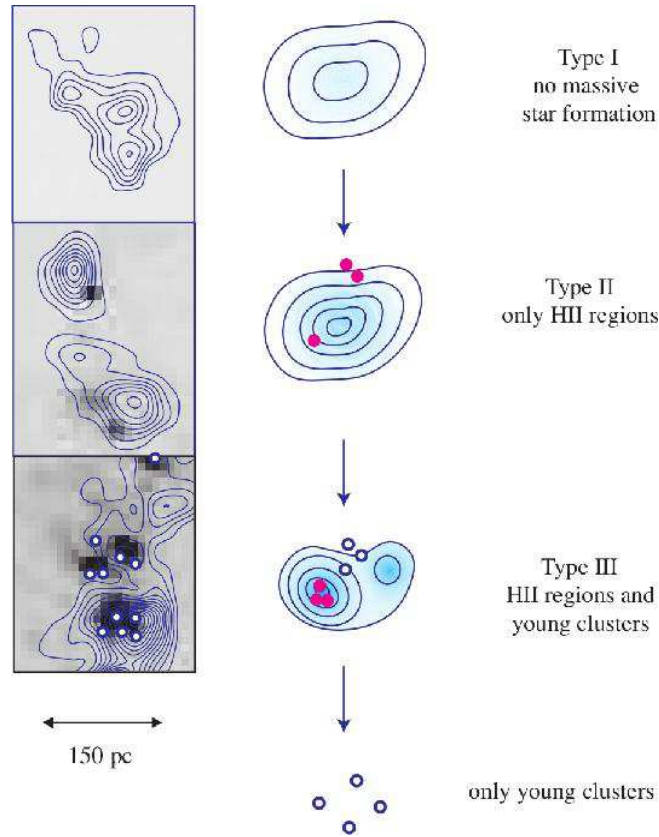}
\caption{Evolutionary sequence of molecular clouds from Kawamura et al. (2009).
\label{fig_3}}
%\end{minipage}
\end{figure}

\section{Energy Feedback and Interstellar Shells}

To study the energy feedback from massive stars, one ought to bear in mind 
that both stars and the ISM evolve, and that their evolutions are always
intertwined, especially for a system like an OB association or a cluster.
From birth to a few Myr old, massive O stars photo-ionize the ambient ISM
to form an H\,{\sc ii} region.  Starting at $\sim$5 Myr, O stars start to explode
as supernovae.  The combined action of fast stellar winds and supernova
blasts sweeps the H\,{\sc ii} region into an expanding shell, i.e., a superbubble.
At $\sim$10 Myr, O stars are gone and B stars start to explode as supernovae.
Without ionizing sources, the superbubble recombines and becomes an 
H\,{\sc i} shell.
After $\sim$15 Myr, all massive stars are gone; the remaining lower-mass
stars disperse, and the H\,{\sc i} supershell coasts along.  If the superbubble is
not near dense molecular material, it will have a simple shell structure, such
as N70.  If the superbubble is in a molecular gas-rich environment, the
expanding superbubble can compress the ambient molecular clouds to form more
stars along the shell rim, such as N11 and N44.  Such sequential star formation
can continue for more than 10$^7$ yr to form a supergiant shell reaching 1000 pc
in size.

\begin{figure}[t!]
%\begin{minipage}{13cm}
\centering
\includegraphics[width=14cm]{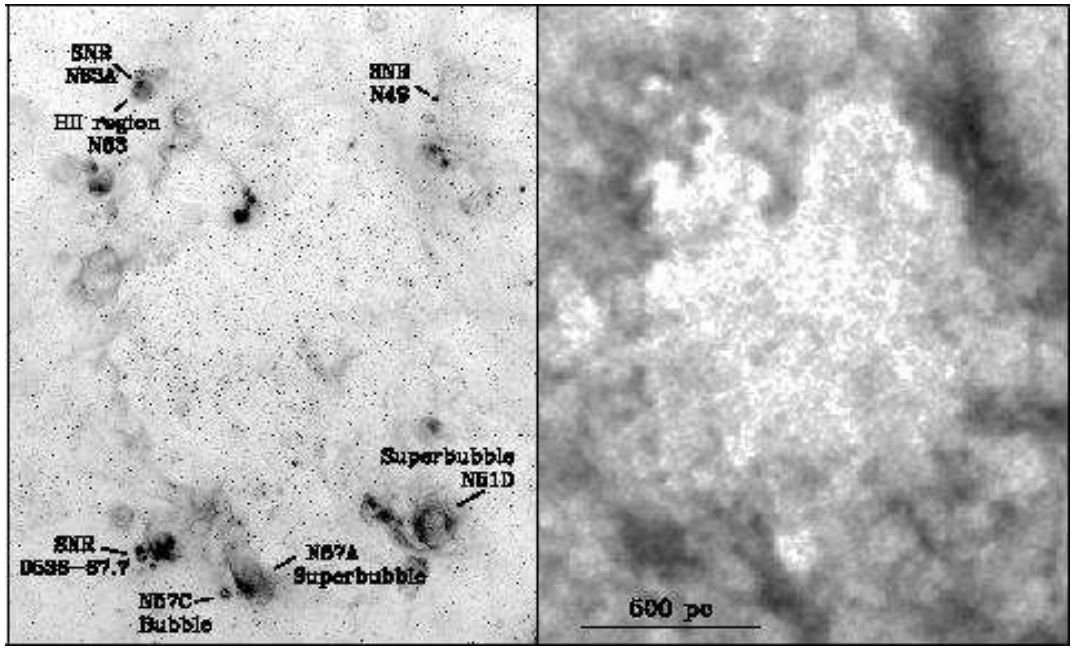}
\caption{H$\alpha$ image (left panel) and H\,{\sc i}
column density map (right panel) of supergiant shell LMC4 in 
the LMC.  The two images have the same field-of-view.
Superbubbles, SNRs, and a bubble are marked in the figure.
\label{fig_4}}
%\end{minipage}
\end{figure}

The most common product of stellar energy feedback is interstellar shells
of various sizes, as illustrated in Figure 4.  An isolated single massive star
with a fast stellar wind can form an interstellar bubble during its main sequence 
stage, a circumstellar bubble (consisting of stellar material previously lost via
a slow wind) during the WR star stage, and a SNR after its final
explosion.  The sizes of bubbles and SNRs can grow up to a few 10's of pc before
dissipating and merging into the ISM.  OB associations can blow superbubbles with 
sizes up to a few hundred pc.  Multiple generations of propagated star formation 
over an extended period of time and space can form supergiant shells of sizes
$\sim10^3$ pc.  The supergiant shell LMC4 in Figure 4 shows that the gas in this 
area has been cleared out, and that on-going star formation takes place along the
supergiant shell rim where dense gas is present.  As massive stars in LMC4 are
concentrated in H\,{\sc ii} regions along the south and northeast rims, the bulk
of gas associated with LMC4 is neutral, as seen in the H\,{\sc i} column density map.

Studies of interstellar shells in recent years have provided answers to many 
puzzles, especially in the seeming lack of visible bubbles and rare detections of
X-ray emission from bubbles.  Massive O stars have fast stellar winds, and if an 
O star is in a reasonably dense ISM, it should blow an interstellar bubble (Weaver 
et al.\ 1977).  However, bubbles are not commonly seen, and the Bubble Nebula is
an exception rather than a rule.  This ``missing bubble'' puzzle has been solved
by studies of the young H\,{\sc ii} regions N11B and N180, where many O 
stars are present 
but no bubbles can be identified morphologically in optical images.  
Using high-dispersion, long-slit echelle spectra of N11B and N180, it has been 
shown that expanding shells are present around O stars, but the expansion 
velocities are only $\sim$20 km~s$^{-1}$.  Such weak shocks cannot compress the 
ambient medium to produce pronounced limb-brightening to be identifiable as a 
bubble (Naz\'e et al.\ 2001).
When the central star evolves off the main sequence and loses its ionizing
power, the bubble and the ambient medium will recombine.  The isothermal 
sound speed of H\,{\sc i} is low, and strong shocks and compression will be produced
by a bubble expanding at 20 km s$^{-1}$; therefore, H\,{\sc i} bubbles are routinely 
detected around massive stars.

The difficulty in detecting diffuse X-ray emission from shocked stellar winds
can be illustrated by the study of the Orion Nebula.  The Orion Nebula hosts an 
O6 star ($\theta^1$ Ori C) with a fast stellar wind, so it is expected to blow a 
bubble and the hot gas in the bubble interior should emit in X-rays.  Diffuse X-ray 
emission from the Orion Nebula was first reported by Ku \& Chanan (1979) using
{\it Einstein} observations, but the diffuse emission was resolved into stars 
by {\it ROSAT} observations (Caillault, Gagne, \& Stauffer 1994).
It was not until 2008 that the diffuse X-ray emission from shocked fast wind 
in the Orion Nebula was truly detected for the first time using
{\it XMM-Newton} observations (G\"udel et al.\ 2008).
The {\it Spitzer} IRAC 8 $\mu$m image of the Orion Nebula region shows
that the Orion cluster has blown a blister-like cavity, 
and the {\it XMM-Newton} observations show diffuse X-ray emission in the 
cavity, at the far end from the cluster.  The plasma temperature determined
from the X-ray spectral fits is $\sim$2$\times$10$^6$~K.
Future searches for diffuse X-ray emission from shocked stellar winds
should bear in mind that IR images may be better at revealing bubble 
cavities in a complex environment and that shocked stellar wind may be 
located far away from the massive stars.

Many circumstellar bubbles blown by WR stars are known, but diffuse X-ray
emission has been detected from only two -- NGC\,6888 and S\,308.
The plasma temperature of NGC\,6888 is $\sim$2$\times$10$^6$~K 
(Wrigge, Wendker, \& Wisotzki 1994) and S\,308 $\sim$1$\times$10$^6$~K
(Chu et al.\ 2003).  X-ray emission from such low plasma temperatures
is extremely soft and the interstellar absorption is high for soft X-rays.
The soft X-ray emission from S\,308 can be detected because it is nearby
and at a high galactic latitude, and thus its foreground absorption 
column density is low.  If S\,308 were in the Galactic plane and
at a larger distance, it would not have been detected.  Therefore,
the higher interstellar absorption is responsible for the low X-ray
detection rate of WR bubbles.

The biggest unanswered question is still the discrepancy between
the observed bubble dynamics and theoretical predictions.
Bubbles are observed to be too small or expand too slowly compared with 
those expected from bubble models using realistic stellar energy input.
X-ray observations of WR bubbles, superbubbles, and planetary nebulae
all show X-ray luminosities much lower than model predictions.  
The clumpiness of stellar winds can reduce the stellar mass loss rate by
a factor of a few, but cannot fully remove the discrepancy between 
observations and model expectations.  Artificially changing the heat
conduction coefficient does not alleviate the discrepancy.  Dynamically
mixing cold nebular material with fast stellar wind (i.e., mass-loading)
may lower the hot gas temperature, and thus raise the cooling rate.
Recent models by Freyer, Hensler, \& Yorke (2003, 2006), Pittard, Dyson,
\& Hartquist (2001), Pittard, Hartquist, \& Dyson (2001), and Arthur (2008)
have grown more sophisticated.  Detailed modelling for a specific bubble
with accurate observations are needed for critical tests of bubble models.

It is worth noting that three cases of nonthermal X-ray emission from
LMC superbubbles have been reported, but two of them are not confirmed by 
more careful analyses.  {\it XMM-Newton} observations of N51D 
(Cooper et al.\ 2004) and {\it Suzaku} observations of N11 (Maddox et 
al.\ 2009) have been reported to show nonthermal diffuse X-ray emission 
from the superbubble interior.  However, Yamaguchi, Sawada, \& Bamba (2010)
have analyzed both {\it XMM-Newton} and {\it Suzaku} observations of N11 
and N51D with a careful background subtraction, and found that neither show 
nonthermal X-ray emission.  They conclude that 30 Dor C is the only LMC
superbubble that shows bona fide nonthermal X-ray emission (Bamba et al.\
2004; Yamaguchi et al.\ 2010).

\section{Acceleration of Cosmic Rays}

Cosmic rays can be accelerated in astronomical shocks (Bell 1978; Bykov 2001).
While observations of some young SNRs have shown evidence of production of cosmic
rays, the origin of Galactic cosmic rays has been largely an unresolved mystery.
The production and diffusion of cosmic rays can be traced by $\gamma$-rays, because
collisions between interstellar protons and cosmic ray protons produce pions and 
each pion decays into two $\gamma$-rays.  It is difficult to associate $\gamma$-ray 
emission with interstellar structures in the Galaxy because of the confusion in 
the Galactic plane.  This is not a problem for the LMC because of its nearly 
face-on orientation.

Recently the {\it Fermi Gamma-Ray Space Telescope} detected $\gamma$-ray 
emission from the LMC, and 
provided the first spatially-resolved global view of $\gamma$-rays from a 
nearly face-on galaxy.  Analyses of the first year's {\it Fermi} LAT observations 
of the LMC find (1) the brightest $\gamma$-ray emission appears
to be centered near the 30 Dor giant H\,{\sc ii} region, but not 
its central R136 cluster; (2) fainter $\gamma$-ray emission 
is detected in the northern part of the LMC; (3) $\gamma$-ray 
emission shows little correlation with the total column
density of the interstellar gas; and (4) $\gamma$-ray emission 
appears to be confined to massive star forming regions.  
These findings indicate that cosmic rays are accelerated 
in massive star forming regions and that the diffusion length of GeV 
cosmic ray protons is relatively short (Abdo et al.\ 2010).

The distribution of $\gamma$-ray emission can be compared with the 
underlying stellar and interstellar components.  We have extracted 
the contours from the integrated $>$100 MeV emissivity map of the LMC 
(the left panel of Figure 10 of Abdo et al.\ 2010), assuming that all 
$\gamma$-rays are diffuse in origin, and plotted these 
contours in Figure 5 over the H$\alpha$ image, H\,{\sc i} column density 
map (Kim et al.\ 2003), {\it ROSAT} PSPC X-ray mosaic (made by 
S. Snowden), and star formation rates 6.3, 12.5, 
and 25 Myr ago (Harris \& Zaritsky 2009).

Figure 5 shows that the $>$100 MeV emissivity is well 
correlated with the star formation within the last 
6--12 Myr not only for the two $\gamma$-ray peaks, 
but also for the faint extension to the west.
No correlation is seen for star formation at 25 Myr or earlier.
Massive-star progenitors of supernovae have a 
lifetime ranging from a few to $\sim$15 Myr; thus, the 
correlation of $\gamma$-ray emissivity with the 6.3 and 
12.5 Myr star formation rates indicates that supernovae play a 
major role in the acceleration of cosmic rays.

Less than 10\% of massive stars are formed in isolation
(Zinnecker \& Yorke 2007) and produce classical SNRs.  
The great majority of massive stars are formed 
in groups, such as OB associations and clusters in single
bursts of star formation, or star clouds in propagated star
formation.  OB associations and clusters a few Myr old produce
superbubbles up to $\sim$200 pc across, while propagated
star formation over 10$^7$ yr produce supergiant shells
$\sim$1000 pc in size (Chu 2008).  The H$\alpha$ image 
and H\,{\sc i} column density map of the LMC (Figure 5) indeed show 
superbubbles and 
supergiant shells in areas where the star formation rate was
high within the last $\sim$12 Myr. Therefore, the $\gamma$-ray
emission is also well correlated with superbubbles and 
supergiant shells.  Within the brightest $\gamma$-ray peak,
many superbubbles exist in the 30 Dor region; the second
brightest $\gamma$-ray peak corresponds to the supergiant shell
LMC4 whose periphery is dotted with superbubbles; and the
faint western extension of diffuse $\gamma$-rays corresponds
to the supergiant shell LMC8 (cataloged by Meaburn 1980).

It is not possible that uncataloged point sources, such as pulsars, 
contribute to the diffuse emission because the $\gamma$-ray emission 
is well correlated with the diffuse soft X-ray emission (Figure 5c) 
and {\it ROSAT} HRI observations have demonstrated that the diffuse 
X-ray emission is truly diffuse (Chu \& Snowden 1998).
Therefore, it may be concluded that the collective, interacting SNR 
shocks within superbubbles and supergiant shells produced by massive
stars formed in the last $\sim$15 Myr have accelerated the cosmic rays in 
the LMC that are responsible for the $>$100 MeV $\gamma$-rays detected 
by {\it Fermi}. 

\begin{figure}[h!]
% \vspace*{-2.0 cm}
\begin{center}
 \includegraphics[width=5.5in]{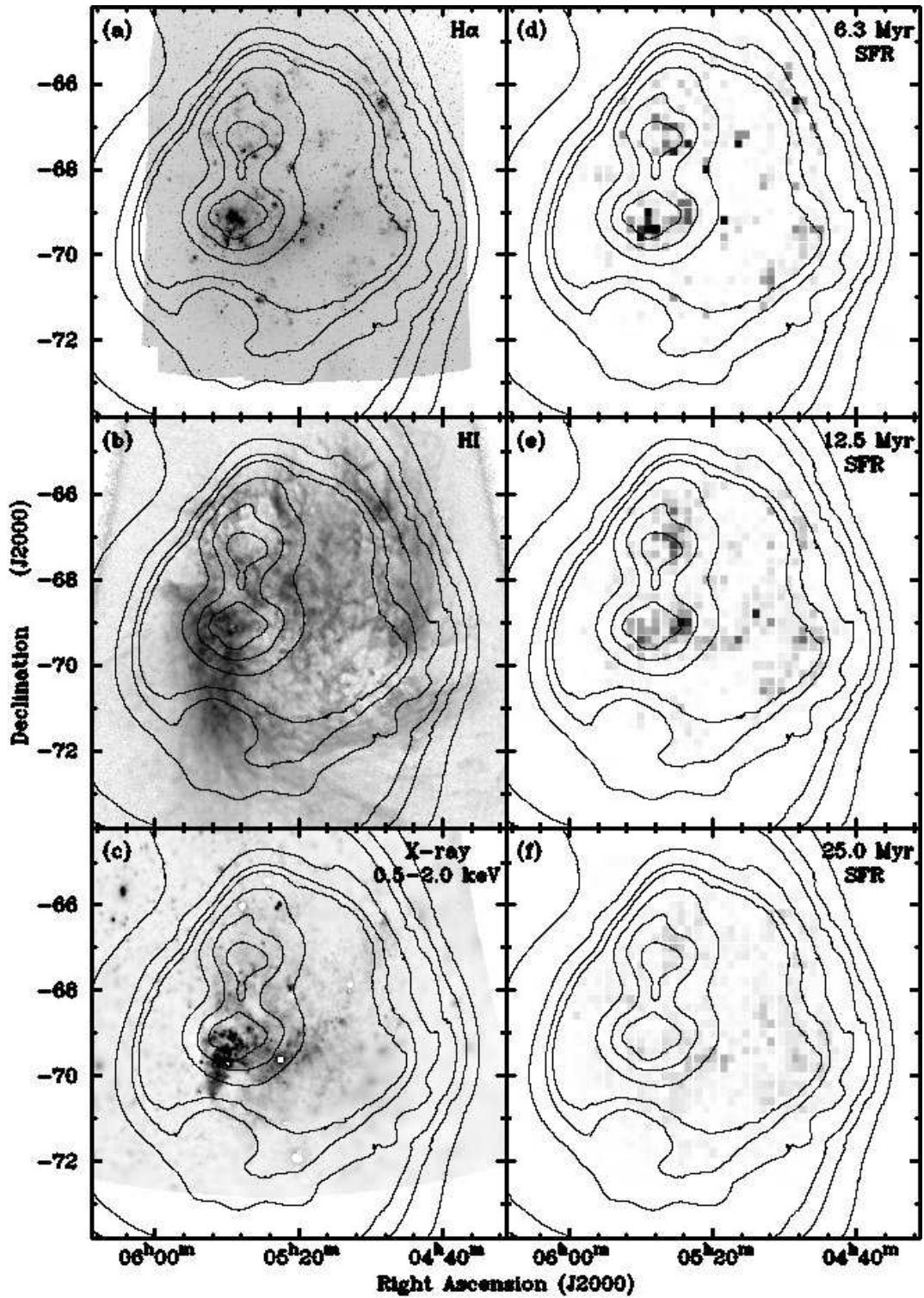}
% \vspace*{-1.0 cm}
 \caption{{\it Fermi} integrated $>$100 MeV emissivity contours
overplotted on (a) MCELS H$\alpha$ image, (b) H\,{\sc i} column density map 
from Kim et al.\ (2003), (c) {\it ROSAT} PSPC mosaic in 0.5-2.0 keV 
made by S. Snowden, (d-f) maps of star formation rate at ages of 
6.3, 12.5, and 25 Myr from Harris \& Zaritsky (2009). 
The H$\alpha$, H\,{\sc i}, and {\it ROSAT} X-ray images have been re-cast 
to the same projection scheme as the {\it Fermi} emissivity map and 
the star formation rate maps.  The 30 Dor giant H\,{\sc ii} region is near 
the brightest $\gamma$-ray peak.
}   \label{fig5}
\end{center}
\end{figure}

\begin{figure}[th!]
% \vspace*{-2.0 cm}
\begin{center}
 \includegraphics[width=5.0in]{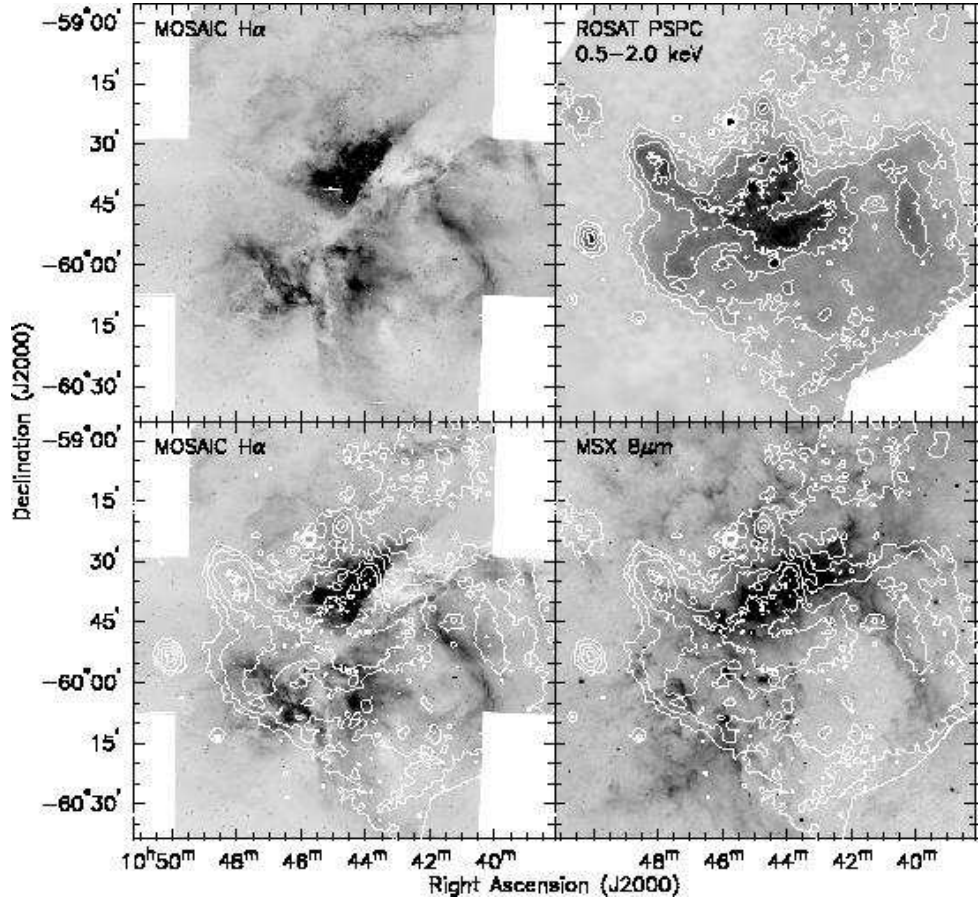}
% \vspace*{-1.0 cm}
 \caption{The top left panel shows the H$\alpha$ emission from the Carina Nebula 
imaged using the MOSAIC II camera on the CTIO Blanco 4m.  The 
top right panel shows the X-ray emission in the 0.5-2.0 keV band as 
seen by the {\it ROSAT} PSPC (courtesy S. Snowden).  The bottom left panel
shows the {\it ROSAT} X-ray emission contours overlaid on the H$\alpha$ image,
while the bottom right panel shows the X-ray contours overlaid on 
the {\it MSX} observations at 8$\mu$m.  
}   \label{fig6}
\end{center}
\end{figure}

\section{Anatomy of the Carina Nebula}

The Carina Nebula, at $\sim$2~kpc, is the nearest, most unobscured,
giant H\,{\sc ii} region in our Galaxy.  As a site of active star formation, 
it hosts the highest concentration of the earliest O stars in the Galaxy as
well as one of the most massive luminous blue variable, $\eta$ Car
(e.g., Walborn 1971; Walborn et al.\ 2002).  As shown in Figure 6,
the H$\alpha$ image of the Carina Nebula reveals non-uniform extinction
over the face of the H\,{\sc ii} region, particularly the V-shaped dust lane 
running through its waist.  

Ever since the first detection of X-ray emission by the {\it Einstein X-ray Observatory}
(Seward \& Chlebowski 1982), the nature of the energy source has been debated.
An O5 star with $L_{\rm w} \sim 10^{35}$  ergs s$^{-1}$ injects 2$\times$10$^{49}$ 
ergs into the ambient ISM during its 5 Myr lifetime, and a WR star can inject
2$\times$10$^{50}$ ergs during 0.5 Myr.  A supernova can deposit 
10$^{50}$ -- 10$^{52}$ ergs of explosion energy into the ISM.
Thus, both fast stellar winds and supernovae could be important energy sources
for the ISM in the Carina Nebula.
Based on the absence of nonthermal radio emission and the presence of massive stars
with powerful stellar winds, Seward \& Chlebowski (1982) suggested that the hot, 
X-ray-emitting gas is powered by the stellar winds.  The only evidence 
for recent supernovae are an anomalously high column density ratio of N(Mn)/N(Fe) 
observed toward  HD\,93205 in the Tr\,16 cluster at the heart of the Carina Nebula 
(Laurent et al.\ 1982) and the recent discovery of a neutron star candidate in the
region (Hamaguchi et al.\ 2009). 

The {\it ROSAT} PSPC X-ray image of the Carina Nebula (Figure 6) shows diffuse
emission in the vicinity of the cluster Tr16, roughly along the V-shaped
dust lane, and extending to the southwest.  This diffuse X-ray emission is
confirmed by deeper, higher-resolution {\it Chandra} ACIS-I observations in 
a recent Msec campaign (Townsley et al.\ 2010). 
The southwest extension of the diffuse X-ray emission is coincident with
a superbubble revealed by the {\it MSX} 8~$\mu$m image (Figure 6).
The 8 $\mu$m emission detected by {\it MSX} suffers much less from the local 
extinction; thus the {\it MSX} 8 $\mu$m image can reveal the true underlying
interstellar structure of the Carina Nebula.  A second supershell to the northeast
of Tr16 is also seen in the 8 $\mu$m image, but diffuse X-ray emission is 
detected only near its base where the brightest H$\alpha$ emission is seen.

\begin{figure}[b!]
\begin{center}
\includegraphics[width=4.5in]{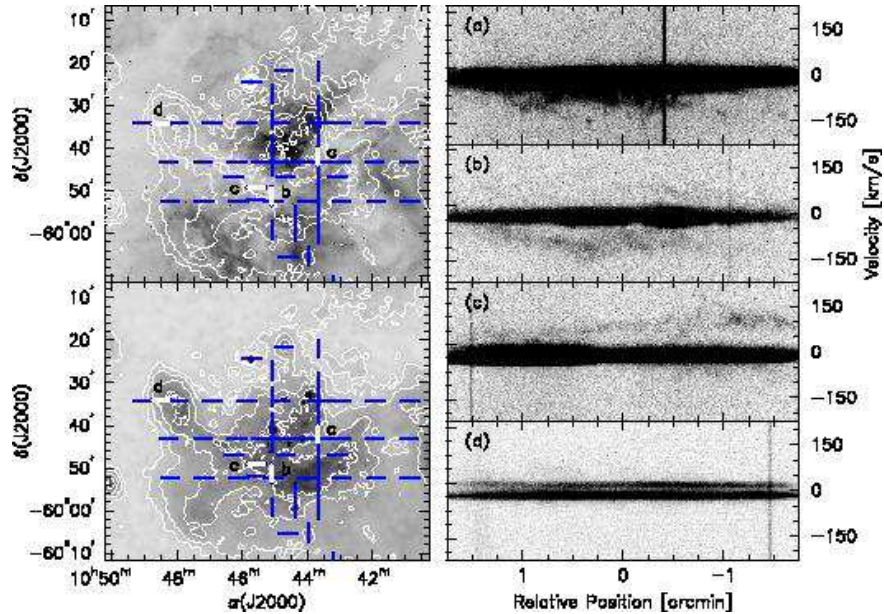}
\vspace{0.8cm}
\caption{(Top Left) An H$\alpha$ image of the central portion of the Carina Nebula 
overlaid with contours to show the X-ray emission.  The horizontal and 
vertical marks show the positions of our echelle spectra.  (Bottom left) 
The {\it ROSAT} PSPC X-ray image in the 0.5--2.0~keV energy band with the 
same slit positions overlaid.  (Right) Each panel shows an echellogram 
for the [N\,{\sc ii}] $\lambda$6584 emission line.  The position of each 
long-slit is highlighted and marked in the panels to the left. 
\label{fig_7}}
\end{center}
\end{figure}

Recently, we have obtained high-dispersion spectroscopic observations of the
H$\alpha$ and [N\,{\sc ii}] lines at positions 
throughout the Carina Nebula (Figure 7). These observations show line-splitting 
indicating expansion velocities of $\sim$15--30~km~s$^{-1}$ at positions that 
correspond to the {\it MSX} cavities and centered on Tr\,16.  Higher-velocity 
components with typical blue-shifted velocity offsets up to $-$180~km~s$^{-1}$, 
and in a few cases red-shifted velocity offsets up to +130~km~s$^{-1}$ are 
observed at some positions in and around the dust lane that bisects the Carina
Nebula.  The higher velocity components appear to be associated with faint 
``frothy'' emission from clumps with linear sizes as small as 0.05~pc extending 
over areas up to 1-3~pc.  These higher velocity components may result from 
wind-ablated material from the dense gas along the waist of Carina.  
A more detailed analysis will be presented in Gruendl et al.\ (in preparation).

%\begin{figure}[h]
%\begin{minipage}{8cm}
%\centering
%\includegraphics[width=8cm]{fig2.pdf}
%\caption{This is the caption of the left-hand side figure. \label{fig_2}}
%\end{minipage}
%\hfill
%\begin{minipage}{8cm}
%\centering
%\includegraphics[width=8cm]{fig3.jpg}
%\caption{This is the caption of the right-hand side figure. \label{fig_3}}
%\end{minipage}
%\end{figure}

%
% USE A SECTION WITHOUT NUMBER FOR THE ACKNOWLEDGEMENTS
%
\section*{Acknowledgements}
YHC would like to thank the generous support of the meeting organizer.  Some of
our research work has been supported by a number of NASA grants through JPL and
STScI and an NSF grant.
%
% BEGIN THE REFERENCE LIST WITH \beginrefer
% USE \refer BEFORE THE REFERENCES AND BEGIN A NEW PARAGRAPH AFTER THE 
% REFERENCE !
% DO NOT FORGET TO END THE LIST WITH \endrefer
%
\footnotesize
\beginrefer

\refer Abdo, A.\ A., Ackermann, M., Ajello, M., et al.\ 2010, A\&A, 512, 7

\refer Arthur, S.\ J.\ 2008, IAUS, 250, 355

\refer Bamba, A., Ueno, M., Nakajima, H., \& Koyama, K.\ 2004, ApJ, 602, 257 

\refer Bell, A.~R.\ 1978, MNRAS, 182, 147 

\refer Book, L.\ G., Chu, Y.-H., Gruendl, R.\ A., \& Fukui, Y.\ 2009, AJ, 137, 3599

\refer Bykov, A.~M.\ 2001, Space Science Reviews, 99, 317

\refer Caillault, J.-P., Gagne, M., \& Stauffer, J.~R.\ 1994, ApJ, 432, 386 

\refer Chen, C.-H.\ R., Chu, Y.-H., Gruendl, R.\ A., Gordon, K.\ D.,
       \& Heitsch, F.\ 2009, ApJ, 695, 511

\refer Chen, C.-H.\ R., Indebetouw, R., Chu, Y.-H., et al.\ 2010, ApJ, 721, 1206

\refer Chu, Y.-H.\ 2008, IAUS, 250, 341 

\refer Chu, Y.-H., Guerrero, M.~A., Gruendl, R.~A., Garc{\'{\i}}a-Segura, G., 
       \& Wendker, H.~J.\ 2003, ApJ, 599, 1189 

\refer Chu, Y.-H., \& Snowden, S.\ L.\ 1998, AN, 319, 101

\refer Cooper, R.\ L., Guerrero, M.\ A., Chu, Y.-H., Chem, C.-H.\ R., 
        Dunne, B.\ C.\ 2004, ApJ, 605, 751

\refer Freyer, T., Hensler, G., \& Yorke, H.~W.\ 2003, ApJ, 594, 888 

\refer Freyer, T., Hensler, G., \& Yorke, H.~W.\ 2006, ApJ, 638, 262

\refer Fukui, Y., Kawamura, A., Minamidani, T., et al.\ 2008, ApJS, 178, 56

\refer Gruendl R.~A. \& Chu Y.-H.\ 2009, ApJS, 184, 172

\refer G{\"u}del, M., Briggs, K.~R., Montmerle, T., Audard, M., Rebull, L., 
       \& Skinner, S.~L.\ 2008, Science, 319, 309 

\refer Hamaguchi, K., Corcoran, M.\ F., Ezoe, Y., et al.\ 2009, ApJ, 695, L4

\refer Harris, J., \& Zaritsky, D.\ 2009, AJ, 138, 1243

\refer Hughes, A., Wong, T., Ott, J., et al.\ 2010, MNRAS, 406, 2065

\refer Kawamura, A., Mizuno, Y., Minamidani, T., et al.\ 2009, ApJS, 184, 1 

\refer Kim, S., Staveley-Smith, L., Dopita, M.\ A.\ 2003, ApJS, 148, 473

\refer Ku, W.~H.-M., \& Chanan, G.~A.\ 1979, ApJ, 234, L59 

\refer Laurent, C., Paul, J.~A., \& Pettini, M.\ 1982, ApJ, 260, 163

\refer Leitherer, C., Schaerer, D., Goldader, J.\ D.\ 1999, ApJS, 123, 3

\refer Maddox, L.\ A., Williams, R.\ M., Dunne, B.\ C., \& Chu, Y.-H.\
       2009, ApJ, 699, 911

\refer Meaburn, J.\ 1980, MNRAS, 192, 365

\refer Meixner, M., Gordon, K.\ D., Indebetouw, R.\ 2006, AJ, 132, 2268

\refer Naz{\'e}, Y., Chu, Y.-H., Points, S.~D., Danforth, C.~W., Rosado, M., 
       \& Chen, C.-H.~R.\ 2001, AJ, 122, 921 

\refer Pittard, J.~M., Dyson, J.~E., \& Hartquist, T.~W.\ 2001, A\&A, 367, 1000

\refer Pittard, J.~M., Hartquist, T.~W., \& Dyson, J.~E.\ 2001, A\&A, 373, 1043

\refer Puerari, I., Block, D.\ L., Elmegreen, B.\ G.\, \&
       Bournaud, F.\ 2010, presented at ``Galaxies and their Masks'' 
       (2010arXiv1008.1044P)

\refer Seale, J.~P., Looney, L.~W., Chu, Y.-H., et al.\ 2009, ApJ, 699, 150 

\refer Seward, F.~D., \& Chlebowski, T.\ 1982, ApJ, 256, 530

\refer Smith, R.\ C., \& the MCELS Team 1999, IAUS, 190, 28

\refer Townsley, L.\ K., Broos, P.\ S., Chu, Y.-H., et al.\ 2010, ApJS, submitted

\refer Vaidya, K., Chu, Y.-H., Gruendl, R.\ A., Chen, C.-H.\ R., \&
        Looney, L.\ W.\ 2009, ApJ, 707, 1417

\refer Walborn, N.~R.\ 1971, ApJ, 167, L31

\refer Walborn, N.~R., Danks, A.~C., Vieira, G., \& Landsman, W.~B.\ 2002, ApJS, 140, 407

\refer Weaver, R., McCray, R., Castor, J., Shapiro, P., \& Moore, R.\ 1977, ApJ, 218, 377

\refer Wrigge, M., Wendker, H.~J., \& Wisotzki, L.\ 1994, A\&A, 286, 219

\refer Yamaguchi, H., Sawada, M., \& Bamba, A.\ 2010, ApJ, 715, 412 

\refer Yang, C.-C., Gruendl, R.A., Chu, Y.-H., Mac Low, M.-M.,
       \& Fukui, Y.\ 2007, ApJ, 671, 374

\refer Zaritsky, D., Harris, J., Thompson, I.~B., \& Grebel, E.\ K.\ 2004, AJ, 128, 1606

\refer Zinnecker, H., \& Yorke, H.\ W.\ 2007, ARAA, 45, 481

\endrefer           
\end{document}